\documentclass[12pt]{article}

\usepackage[utf8]{inputenc}
\usepackage{import}
\usepackage[toc,page]{appendix}
\usepackage{latexsym,amsfonts,amsmath,amssymb,mathrsfs,bbold,mathtools,esint,amsthm,mathtools,mathptmx}
\usepackage{dsfont}
\usepackage{mathtools}
\usepackage{float}
\usepackage{graphicx}
\usepackage{cite}
\usepackage{hyperref}
\usepackage{setspace}
\usepackage{color}
\usepackage{cleveref}
\numberwithin{equation}{section}
\usepackage[colorinlistoftodos]{todonotes}
\usepackage[affil-it]{authblk}
\usepackage{float}
\usepackage{indentfirst}
\usepackage{soul}
\usepackage{booktabs}
\usepackage{eso-pic,graphicx}
\usepackage{nicefrac}
\usepackage{epsfig}

\usepackage[a4paper]{geometry}
\usepackage{a4wide}

\usepackage{color,hyperref}
\definecolor{darkblue}{rgb}{0.0,0.0,0.3}
\hypersetup{colorlinks,breaklinks,
            linkcolor=darkblue,urlcolor=darkblue,
            anchorcolor=darkblue,citecolor=darkblue}

\title{On quantum Freidel-Maillet algebra for non-ultralocal integrable systems}
\author[1]{A. Melikyan\footnote{\href{mailto:amelik@gmail.com}{amelik@gmail.com}}}
\author[2]{G. Weber\footnote{\href{mailto:gabrielweber@usp.br}{gabrielweber@usp.br}} }
\affil[1]{Instituto de Física\\
Universidade de Brasília\\
70910-900, Brasília, DF, Brasil}
\affil[2]{Escola de Engenharia de Lorena\\
Universidade de São Paulo\\
C.P. 116, 12600-970, Lorena, SP, Brasil}

\begin{document}
\maketitle
\begin{abstract}
	We consider the quantum algebra of transition matrices for non-ultralocal integrable systems, and show that a regularization of the singular operator products in the quantum algebra via Sklyanin's product leads to well-defined expressions, reproducing in the classical limit Maillet's symmetrization prescription for Poisson brackets.
\end{abstract}

\newpage
\section{Introduction}

The quantization of non-ultralocal integrable systems is a challenging open problem, which has become especially relevant and interesting since the discovery that $AdS_{5} \times S^{5}$ string theory is a classically integrable system of this type (for a review see \cite{Beisert:2010jr} and references therein). Out of several existing approaches, Maillet's approach \cite{Maillet:1985ek,Freidel:1991jx,Freidel:1991jv} seems to be the simplest and most systematic in order to construct the action-angle variables, and understand the classical integrability. This is also true for higher order non-ultralocal theories, i.e., the theories for which the algebra of the $L$-operators contains also the second derivatives of the delta-function. Such cases  appear, for example, in the $su(1|1)$ sector of the $AdS_{5} \times S^{5}$ string theory (see for details \cite{Melikyan:2012kj,Melikyan:2014yma}), where it can be shown that the algebra of the transition matrices has the same form as in the case considered by Maillet, with some appropriate shift in the $(r,s)$-pair. 

The fundamental construction underlying Maillet's method is the symmetrization prescription for the Poisson brackets and the corresponding generalization for the nested Poisson brackets. To introduce Maillet's symmetrization procedure one considers an $n$-nested Poisson bracket for transition matrices $T(x_{i},y_{i};\lambda_{i})$:
\begin{align}
\Delta^{n}(x_{i},y_{i};\lambda_{i}) = \left\{ T(x_{1},y_{1};\lambda_{1}) \overset{\otimes}{,} \left\{ \ldots\overset{\otimes}{,}\left\{ T(x_{n},y_{n};\lambda_{n}) \overset{\otimes}{,} \;T(x_{n+1},y_{n+1};\lambda_{n}) \right\} \ldots \right\} \right\}, \label{intro:nested_brackets}
\end{align}
and for any subset of $l=p+q$ coinciding points $x_{\alpha_{1}}= \ldots =x_{\alpha_p}=y_{\beta_{1}}= \ldots = y_{\beta_q}=z$, one defines the left-hand side of \eqref{intro:nested_brackets} by: %by the symmetrization of Poisson brackets, defined as follows:
\begin{align}
\Delta^{n}(z;\lambda_{i}) := \lim_{\epsilon \rightarrow 0} \frac{1}{l!} \sum_{\sigma  \,  {\scriptscriptstyle\in} \, \mathds{P}} \Delta^{n} \left(x_{\alpha_1} + \epsilon \sigma(1),\ldots,y_{\beta{q}} + \epsilon \sigma(l);\lambda_{i} \right), \label{intro:symmetrization}
\end{align}
where for simplicity of notations we omitted in $\Delta^{n}(x_{i},y_{i};\lambda_{i})$ the dependence on the coordinates different from $z$, and the symbol $\mathds{P}$ indicates the sum over all possible permutations of $(1,\ldots,l)$. For example, the symmetrization procedure yields:
\begin{align}
	\{&T(x,y;\lambda) \overset{\otimes}{,} \;T(x,y';\mu) \}_{{M}} \nonumber\\
	& = \frac{1}{2} \lim_{\epsilon \rightarrow 0}  \left(\{T(x - \epsilon,y;\lambda) \overset{\otimes}{,} \;T(x + \epsilon,y';\mu) \} + \{T(x + \epsilon,y;\lambda) \overset{\otimes}{,} \;T(x - \epsilon,y';\mu) \} \right),\label{intro:PB_sym}
\end{align}
where the subscript ${M}$ simply indicates the symmetrized as above Poisson bracket.

This prescription was shown to be consistent with the Jacobi identities following from the classical algebra of transition matrices. We recall that the classical Maillet algebra \cite{Maillet:1985ek,Freidel:1991jx,Freidel:1991jv} for transition matrices, corresponding to equal and adjacent intervals with $x>y>z$, can be written in the following form:
\begin{align}
		\{T_{1}(x,y;\lambda), \; T_{2}(x,y;\mu) \}_{{M}}
		&= a_{12}(\lambda,\mu) \; T_{1}(x,y;\lambda)T_{2}(x,y;\mu)
		- T_{1}(x,y;\lambda)T_{2}(x,y;\mu) \;d_{12}(\lambda,\mu), \notag \\
		\{T_{1}(x,y;\lambda), \;T_{2}(y,z;\mu) \}_{{M}} &= T_{1}(x,y;\lambda) \; b_{12}(\lambda,\mu) \; T_{2}(y,z;\mu). \label{sp:Maillet_classical_algebra}
\end{align}
Here, we restrict our analysis to a simpler case of \cite{Freidel:1991jv} involving only bosonic fields, and for which the matrices $a_{12}$, $d_{12}$ and $b_{12}=c_{21}$ depend only on the spectral parameters. It was shown in \cite{Freidel:1991jx,Freidel:1991jv} that the Jacobi identiy implies the following Yang-Baxter-like constraints on the $(abcd)$-algebra \eqref{sp:Maillet_classical_algebra}:
\begin{align}
	\left[ a_{12}(\lambda,\mu), a_{13}(\lambda,\nu) \right] + \left[ a_{12}(\lambda,\mu), a_{23}(\mu,\nu) \right] + \left[ a_{13}(\lambda,\nu), a_{23}(\mu,\nu) \right] &= 0,\label{sp:YB_a} \\
\left[ d_{12}(\lambda,\mu), d_{13}(\lambda,\nu) \right] + \left[ d_{12}(\lambda,\mu), d_{23}(\mu,\nu) \right] + \left[ d_{13}(\lambda,\nu), d_{23}(\mu,\nu) \right] &= 0,\label{sp:YB_d}\\
\left[ b_{12}(\lambda,\mu), d_{13}(\lambda,\nu)\right] + \left[b_{32}(\nu,\mu), d_{13}(\lambda,\nu)\right] + \left[ b_{32}(\nu,\mu),b_{12}(\lambda,\mu) \right] &=0,\label{sp:YB_bd}\\
\left[ a_{32}(\nu,\mu), c_{21}(\mu,\lambda)\right] + \left[a_{32}(\nu,\mu), c_{31}(\nu,\lambda)\right] + \left[ c_{31}(\nu,\lambda),c_{21}(\mu,\lambda) \right] &=0.\label{sp:YB_ac}
\end{align}

A general derivation of Maillet's symmetrization procedure from first principles has so far been missing. A related problem is the construction of a quantum algebra corresponding to \eqref{sp:Maillet_classical_algebra} which in the classical limit reproduces Maillet's symmetrization procedure, and the classical Jacobi identities \eqref{sp:YB_a}-\eqref{sp:YB_ac}. The principal difficulty lies in the fact that the standard quantum commutators cannot naively reproduce the symmetrized Poisson brackets \eqref{intro:PB_sym} in the classical limit.

In this short note we propose a quantum algebra for non-ultralocal integrable systems which does reproduce Maillet's symmetrized Poisson brackets in the classical limit. We stress that this is done directly in the continuous case, without appealing to any lattice formulation of the theory. The key idea is based on the fact that the quantum fields, together with the quantum transition matrices and the operators defining the algebraic relations, should be treated as operator-valued distributions. Therefore, the quantum algebra of transition matrices contains singularities due to operator products at the same point, and in order to avoid these singularities one should regularize such operator products. The classical theory is then obtained from this regularized quantum theory, and reproduces Maillet's symmetrized Poisson bracket prescription in a natural manner. In addition, the classical Jacobi identities \eqref{sp:YB_a}-\eqref{sp:YB_ac} are found from the corresponding quantum identities involving only usual commutators. 

Our paper is organised as follows: in section \ref{sec:SP_JI}, we discuss how to address the problem of ill-defined operator products when formulating a continuous quantum algebra by employing Sklyanin product. In section \ref{sec:QA}, we propose a quantum algebra of transition matrices, which in the classical limit reproduces the symmetric prescription of Maillet for Poisson brackets. Finally, in section \ref{sec:conclusion}, we summarise our results and point out some interesting directions and open problems.

\section{Sklyanin's product and quantum Jacobi identity} \label{sec:SP_JI}

To formulate a well-defined algebra for quantum transition matrices one must first deal with the singularities associated with operator products at the same point. The most natural way to solve this problem is to resort to the methods of quantum field theory where such singularities are dealt with by means of renormalisation techniques. The latter, in turn, can be well defined and formulated in terms of quantum fields treated as operator-valued distributions. This corresponds to ``smearing'' the fields over some small region about the point of the product, i.e,  one considers the fundamental quantum field to be:
\begin{equation}
	\phi_{f}(x) = \int \phi(y)f(x-y) dy,\label{sp:field_distribution}
\end{equation}
where $f(x)$ is an element in the Schwartz space of test functions. This is strictly formulated in the framework of axiomatic quantum field theory (see, for example, the monograph \cite{streater2000pct}). 

The key obstacle in formulating a well-defined quantum algebra for transition matrices, corresponding to the classical expressions \eqref{sp:Maillet_classical_algebra}, is Schwartz's theorem on the impossibility of defining a product of two distributions with natural properties (for an overview, see, for example, \cite{Zeidler:2006rw,Zeidler:2009zz}). Namely, it is impossible to define a product of distributions which satisfies linearity, distributivity, commutativity, associativity and the Leibniz rule. Although it is possible to define a product of distributions in some exceptional cases, for instance when their singular supports are disjoint, in general one must look for extensions of Schwartz's distribution theory, e.g., the Colombeau algebras of generalised functions \cite{colombeau1990,colombeau2000new}, where it is possible to define such products (for some applications in physics, see \cite{grosser2001geometric} and references therein). The microlocal analysis, based on the concept of wavefront sets, is another tool that has been used in the context of quantum field theory to provide useful criteria for properly defining the product of distributions (for a review, see \cite{brouder:2014jp} and references therein). Another interesting possibility is the theory of Sato's hyperfunctions \cite{sato:1959av,sato:1969bb}, which also contains Schwartz's distribution theory and relies on the boundary behaviour of analytic  functions.

We postpone the discussion of all these nuances to a future publication, and consider here a preliminary treatment of regularising the operator product at the same point by means of the \emph{Sklyanin product} \cite{Sklyanin:1988}, which again corresponds to a ``smearing'' of the product of operators around the singularity region. More precisely, for two operators $A(x)$ and $B(x)$, it is defined as follows:
\begin{equation}
A(x) \circ B(x) 
\equiv  \lim_{\Delta/2 \to \epsilon} \frac{1}{(\Delta - \epsilon)^{2}} \fint\limits_{\Delta S_{1} \cup \Delta S_{2}}d\zeta d\xi A(\zeta)B(\xi). \label{sp:sklyanin_prod}
\end{equation}
The notation $\fint\limits_{\Delta S_{1} \cup \Delta S_{2}}$ means that the integration is taken over a square of side $\Delta$, minus a strip of width $\epsilon$ around the diagonal $\zeta=\xi$, and the areas $\Delta S_{1}$ and $\Delta S_{2}$ correspond to the regions above the line $\zeta=\xi + \epsilon$ and below the line $\zeta = \xi - \epsilon$. This essentially means that we ``smear'' the product of two operators around an arbitrary small area of size $\Delta$, avoiding the singularity at $\zeta=\xi$. The parameter $\epsilon$ is the regularisation parameter of the theory, and should be taken to zero only at the end of all computations. 

It is clear from \eqref{sp:sklyanin_prod} that if the product of two operators is not singular, then, in the limit $\epsilon \to 0$, it reduces to the usual product. Here we use a slightly more precise version of the definition originally given by Sklyanin in \cite{Sklyanin:1988}, which did not explicitly involve the regularisation parameter $\epsilon$. In other words, we explicitly exclude the entire singular region, parametrised by the length $\epsilon$. We refer to the original paper \cite{Sklyanin:1988} for additional details and properties, and only mention that Sklyanin product was originally introduced in order to write the quantum algebra of the fields of the continuous anisotropic Heisenberg model to ensure the quantum integrability of the model.\footnote{It is worth mentioning that the quantum algebra corresponding to the anisotropic Heisenberg model is not a direct generalisation of the classical algebra, i.e., it is not a Lie algebra, but acquires quadratic quantum corrections, which are written in terms of Sklyanin product. A deeper reason of such algebras is still unclear.} Recently it was also shown that by employing Sklyanin product one can explicitly diagonalize the quantum Hamiltonian for the $su(1|1)$ subsector of strings in $AdS_5 \times S^5$ and obtain the correct $S$-matrix \cite{Melikyan:2014mfa}. In addition, we note that, when discussing Jacobi identities for the fields one has to operate with functionals of the fields, and not with the fields themselves (for a detailed discussion of this issue see \cite{olver1986applications}).\footnote{Strictly speaking one has to still show that the functionals involved satisfy a necessary condition derived in \cite{olver1986applications}.}

For a product of $k$ operators $A_{1}(x), \ldots, A_{k}(x)$, Sklyanin product is defined similarly to \eqref{sp:sklyanin_prod}. The integration should now be performed over the $k$-dimensional cube of side $\Delta$, where all possible singular regions are taken out of the region of the integration. Thus, there are $k!$ integrations over disconnected volume elements of size $\Delta V_{i}$ corresponding to all possible orderings of the variables $\zeta_{1}, \ldots, \zeta_{k}$ separated by the length of the regularisation parameter $\epsilon$. More precisely, we have:
\begin{equation}
A_{1}(x) \circ \ldots  \circ A_{k}(x)
\equiv  \lim_{\Delta/2 \to \epsilon} \frac{1}{\Delta V} \sum\limits_{i=1}^{k!}\int\limits_{\Delta V_{i}} d\zeta_{1} \ldots d\zeta_{k} A_{1}(\zeta_{1}) \cdot \ldots \cdot A_{k}(\zeta_{k}), \label{sp:sklyanin_prod_general}
\end{equation}
where $\Delta V$ is the sum of all disconnected volume elements $\Delta V_{i}$.

It is easy to show that the set of operator-valued functions with product given by \eqref{sp:sklyanin_prod_general} satisfies the following standard relations:
\begin{align}\label{sp:property1}
	\left[ A_1(x) \stackrel{\circ}{,} A_2(x) \circ A_3(x) \right] = A_2(x) \circ \left[ A_1(x) \stackrel{\circ}{,}  A_3(x) \right] + \left[ A_1(x) \stackrel{\circ}{,} A_2(x) \right] \circ A_3(x),  
\end{align}
and
\begin{align}\label{sp:property2}
	\left[ A_1(x) \stackrel{\circ}{,}\left[ A_2(x) \stackrel{\circ}{,} A_3(x) \right] \right] = \left[ A_1(x) \stackrel{\circ}{,} \alpha A_2(x) \circ A_3(x) + A_3(x) \circ A_2(x)  \beta\right],
\end{align}
if
\begin{align}\label{sp:proto_algebra}
	\left[ A_2(x) \stackrel{\circ}{,} A_3(x) \right] = \alpha A_2(x) \circ A_3(x) + A_3(x) \circ A_2(x)  \beta,
\end{align}
where $\alpha, \beta $ are arbitrary constants. Moreover, if the algebra involving only operator-valued functions at different points satisfy the Jacobi identity, the definition \eqref{sp:sklyanin_prod_general} we adopted trivially extends this  Jacobi identity to the case where some arbitrary subset of the points may coincide. In the following, we consider this last point in more detail for the algebra of transition matrices.

Since Sklyanin product \eqref{sp:sklyanin_prod_general} is free of singularities, we can use it to formulate the quantum algebra of transition matrices $T(x,y;\lambda)$. We start from a well-defined Jacobi identity for the case where all points $(u,u',v,v',w,w')$ are different:
\begin{align}
&\left[ T_{1}(u,u';\lambda), \left[ T_{2}(v,v';\mu), T_{3}(w,w';\rho) \right] \right]	+ \mathbb{P}_{13}\mathbb{P}_{23} \left[T_{1}(w,w';\rho), \left[ T_{2}(u,u';\lambda), T_{3}(v,v';\mu) \right] \right] \mathbb{P}_{23}\mathbb{P}_{13}  \nonumber \\
&+ \mathbb{P}_{13}\mathbb{P}_{12} \left[T_{1}(v,v';\mu), \left[ T_{2}(w,w';\rho), T_{3}(u,u';\lambda) \right] \right] \mathbb{P}_{12}\mathbb{P}_{13} =0,\label{sp:JI_all_diff_points}
\end{align}
where $\mathbb{P}$ is the permutation operator acting on the auxiliary spaces. Here, the product between operators is the usual one, because all the points $(u,u',v,v',w,w')$ are different and hence there are no singularities. If, on the other hand some of the points $(u,u',v,v',w,w')$ coincide, then the expression \eqref{sp:JI_all_diff_points} contains a product of operators at the same point, and is, therefore, a singular expression. 

We now formulate a well-defined expression for an arbitrary case of possibly coinciding points. To this end, starting from the well-defined expression \eqref{sp:JI_all_diff_points} and employing Sklyanin product, one obtains the following formula valid for an arbitrary set of points $(x_{1},y_{1},x_{2},y_{2},x_{3},y_{3})$:
\begin{align}
&\left[ T_{1}(x_{1},y_{1};\lambda)\stackrel{\circ}{,}  \left[ T_{2}(x_{2},y_{2};\mu) \stackrel{\circ}{,}  T_{3}(x_{3},y_{3};\rho) \right] \right] \notag \\
&+ \mathbb{P}_{13}\mathbb{P}_{23} \left[T_{1}(x_{3},y_{3};\rho)\stackrel{\circ}{,} \left[ T_{2}(x_{1},y_{1};\lambda)\stackrel{\circ}{,}  T_{3}(x_{2},y_{2};\mu) \right] \right] \mathbb{P}_{23}\mathbb{P}_{13}  \notag \\
&+  \mathbb{P}_{13}\mathbb{P}_{12} \left[T_{1}(x_{2},y_{2};\mu)\stackrel{\circ}{,}  \left[ T_{2}(x_{3},y_{3};\rho)\stackrel{\circ}{,}  T_{3}(x_{1},y_{1};\lambda) \right] \right] \mathbb{P}_{12}\mathbb{P}_{13} = 0.\label{sp:JI_any_points_SP}
\end{align}
It is clear that when all points $(x_{1},y_{1},x_{2},y_{2},x_{3},y_{3})$ are different, the expression \eqref{sp:JI_any_points_SP} trivially reduces to \eqref{sp:JI_all_diff_points}. Note that if there are several sets of coinciding points then Sklyanin product should be taken independently with respect to each set of coinciding points. We emphasise that, although not explicitly indicated, the formula \eqref{sp:JI_any_points_SP} depends on the regularisation parameter $\epsilon$ and, therefore, is a well-defined expression. 

\section{Quantum algebra of transition matrices}\label{sec:QA}

We now turn to our main goal of formulating a quantum algebra between the transition matrices corresponding to the classical algebra \eqref{sp:Maillet_classical_algebra}. As we mentioned earlier, the key problem in writing a quantum algebra is the difficulty to obtain Maillet's symmetrization procedure \eqref{intro:nested_brackets} and \eqref{intro:symmetrization} from the quantum relations in the classical limit. Indeed, it is not obvious how a commutator between two operators, or in general $n$-nested commutators, becomes in the classical limit the symmetrized Poisson brackets \eqref{intro:nested_brackets}. In the classical theory this symmetrization procedure was introduced by Maillet essentially by hand for consistency with the Jacobi identity. 

Here we explain these results by appealing to the quantum theory first, and then considering the classical limit. The main idea, as we have explained above, is that the quantum relations are ill-defined due to the product of operators at the same point, and should be regularised at the very beginning. We have done so above by utilizing Sklyanin product \eqref{sp:sklyanin_prod}, \eqref{sp:sklyanin_prod_general}. To illustrate how Maillet's procedure appears in the classical theory, we show that the commutator of two operator-valued functions goes to the symmetrized Poisson bracket \eqref{intro:PB_sym} in the classical limit. Writing explicitly the commutator between two operator-valued functions $\hat{A}(x)$ and $\hat{B}(x)$ in terms of the definition \eqref{sp:sklyanin_prod}, we obtain:
\begin{equation}
\left[\hat{A}(x)\stackrel{\circ}{,}  \hat{B}(x)\right]
=  \lim_{\Delta/2 \to \epsilon} \frac{1}{(\Delta - \epsilon)^{2}}\left( \iint\limits_{\Delta S_{1}}d\zeta d\xi \left[ \hat{A}(\zeta), \hat{B}(\xi) \right] + \iint\limits_{\Delta S_{2}}d\zeta d\xi \left[ \hat{A}(\zeta), \hat{B}(\xi) \right] \right). \label{QA:sklyanin_prod_commutator}
\end{equation}
In the classical limit \eqref{QA:sklyanin_prod_commutator} becomes:\footnote{Here, C.L. stands for classical limit.}
\begin{equation}
\left[\hat{A}(x)\stackrel{\circ}{,}  \hat{B}(x)\right]_{\substack{C.L.}}
= \lim_{\Delta/2 \to \epsilon} \frac{1}{(\Delta - \epsilon)^{2}}\left( \iint\limits_{\Delta S_{1}}d\zeta d\xi \left\{{A}(\zeta), {B}(\xi) \right\} + \iint\limits_{\Delta S_{2}}d\zeta d\xi \left\{ {A}(\zeta), {B}(\xi) \right\} \right), \label{QA:sklyanin_prod_PB}
\end{equation}
where $A(x)$ and $B(x)$ are already the corresponding classical functions, and $\left\{ A(\zeta), B(\xi) \right\}$ in the right-hand side is the usual Poisson bracket. It is then clear (by invoking the mean value theorem in the regions $\Delta S_{1}$ and $\Delta S_{2}$, where the integrands are smooth functions) that in the limits $\Delta \to \frac{\epsilon}{2}$, followed by the limit $\epsilon \to 0$, the right-hand side of \eqref{QA:sklyanin_prod_PB} reduces to:
\begin{align}
	\left[\hat{A}(x)\stackrel{\circ}{,}  \hat{B}(x)\right]_{\substack{C.L.}}
=  \frac{1}{2}  \lim_{\epsilon \to 0} \left( \left\{ A(x+\epsilon), B(x - \epsilon) \right\} + \left\{ A(x-\epsilon), B(x+\epsilon) \right\} \right).  \label{QA:SP_symmetrization}
\end{align}
Comparing this formula with Maillet's definition of the symmetrized Poisson bracket \eqref{intro:PB_sym} we conclude that:
\begin{align}
	\left[\hat{A}(x)\stackrel{\circ}{,}  \hat{B}(x)\right]_{\substack{C.L.}} = \left\{ A(x),B(x) \right\}_{M}. \label{QA:SP_sym_connection}
\end{align}

This formula shows the connection between Sklyanin product in the quantum case, and Maillet's \emph{ad hoc} construction of symmetrized Poisson brackets. Namely, the classical limit of the commutator, regularised via Sklyanin product, reproduces precisely the symmetrized Poisson bracket. As we discussed above, it is much simpler and more natural to start from the singularities-free quantum algebra, and obtain the classical formulas by taking the corresponding limit of the quantum theory. In this way, Maillet's construction appears naturally. This consideration can also be easily generalised, and the $n$-nested Poisson brackets \eqref{intro:nested_brackets} can be similarly obtained from the general Sklyanin product \eqref{sp:sklyanin_prod_general}.

Using the relation \eqref{QA:SP_sym_connection} one can formulate the quantum algebra of transition matrices directly in the continuous case. To do so, we first recall the lattice algebra for the quantum case proposed by Freidel and Maillet in \cite{Freidel:1991jx,Freidel:1991jv}. It has the following form:
\begin{align}
\hat{A}_{12}T^{(n)}_{1}T^{(n)}_{2} &=T^{(n)}_{2} T^{(n)}_{1}\hat{D}_{12}, \label{QA:lat1}\\
T^{(n)}_{1}T^{(n+1)}_{2} &= T^{(n+1)}_{2} \hat{C}_{12}T^{(n)}_{1},\label{QA:lat2}\\
\left[ T^{(n)}_{1} ,T^{(m)}_{2} \right] &= 0, \quad \text{for  } |n-m| > 1. \label{QA:lat3}
\end{align}
Upon using the quasi-classical expansions $\hat{A}_{12}=1+ i \hbar a_{12}+ \ldots$ for all matrices in \eqref{QA:lat1}-\eqref{QA:lat3}, one obtains the classical lattice algebra:
\begin{align}
	\left\{ T^{(n)}(\lambda) \right. \left.\stackrel{\otimes}{,} T^{(n)}(\mu) \right\} &=  a(\lambda,\mu) \: T^{(n)}(\lambda) \otimes T^{(n)}(\mu) - T^{(n)}(\lambda) \otimes T^{(n)}{}(\mu) \: d(\lambda,\mu), \label{QA:lat1_cl}  \\
	\left\{ T^{(n)}(\lambda) \right. \left.\stackrel{\otimes}{,} T^{(n+1)}(\mu) \right\} &=  - \left( \mathbb{1} \otimes T^{(n+1)}(\mu)  \right) c(\lambda,\mu) \:  \left( T^{(n)}(\lambda) \otimes \mathbb{1} \right) \label{QA:lat2_cl}\\
	\left\{ T^{(n)}(\lambda) \right. \left.\stackrel{\otimes}{,} T^{(m)}(\mu) \right\} &= 0, \quad \text{for  } |n-m| > 1, \label{QA:lat3_cl}
\end{align}
where $T^{(n)}(\lambda) \equiv T(x_{n+1}, x_{n}; \lambda)$ is defined so as to make a connection with the continuous algebra \eqref{sp:Maillet_classical_algebra}.
In passing from the quantum algebra \eqref{QA:lat1}-\eqref{QA:lat3} to the classical one \eqref{QA:lat1_cl}-\eqref{QA:lat3_cl} one obtains, however, the usual Poisson brackets, which are not symmetrized according to Maillet's prescription \eqref{intro:PB_sym} and therefore do not satisfy the classical Jacobi identity. Thus the quantum lattice algebra \eqref{QA:lat1}-\eqref{QA:lat3} cannot reproduce correctly the classical symmetrized Poisson brackets \eqref{sp:Maillet_classical_algebra}.

To solve this problem, we instead propose the following quantum algebra, without resorting to any lattice formulation:
\begin{align}
\hat{A}_{12}T_{1}(x,y;\lambda) {\circ} T_{2}(x,y;\mu) &=T_{2}(x,y;\mu){\circ} T_{1}(x,y;\lambda)\hat{D}_{12}, \label{QA:nolat1_sp}\\
T_{1}(y,z;\lambda) {\circ} T_{2}(x,y;\mu) &= T_{2}(x,y;\mu) {\circ} \hat{C}_{12}T_{1}(y,z;\lambda).\label{QA:nolat2_sp}
\end{align}
These relations are well defined due to Sklyanin product, and upon using as before the quasi-classical expansions for the matrices $(A,B,C,D)$, one obtains:\footnote{We remind that in this paper, we consider only the case $B_{12}=C_{21}$.}
\begin{align}
	\left[ T_1(x,y;\lambda) \stackrel{\circ}{,} T_2(x,y;\mu) \right] & = -i a_{12}(\lambda,\mu) T_1(x,y;\lambda) \circ T_2(x,y;\mu) \label{QA:equal_interval} \\ &+ i T_2(x,y;\mu) \circ T_1(x,y;\lambda) d_{12}(\lambda,\mu) ,\notag \\
	\left[ T_1(y,z;\lambda)\stackrel{\circ}{,}  T_2(x,y;\mu) \right] &= i T_2(x,y;\mu)\circ c_{12}(\lambda,\mu) T_1(y,z;\lambda) \label{QA:adjacent_interval}.
\end{align}

Next, using the quantum algebra \eqref{QA:equal_interval} and \eqref{QA:adjacent_interval} to evaluate the Jacobi identity \eqref{sp:JI_any_points_SP} for all possible combinations of intervals, i.e., equal, adjacent and mixed, we can derive the classical consistency relations \eqref{sp:YB_a}-\eqref{sp:YB_ac} in the classical limit. We also obtain the classical relation $b_{12}(\lambda,\mu) = c_{21}(\mu,\lambda)$ and the antissymetry of the parameters $a_{12}(\lambda,\mu)$ and $d_{12}(\lambda,\mu)$ in the description of the classical algebra under the permutation of the auxiliary spaces corresponding to the spectral parameters $\lambda$ and $\mu$. We note that the properties \eqref{sp:property1} and \eqref{sp:property2} enjoyed by Sklyanin product render the aforementioned calculations a mere repetition of the computation originally performed in \cite{Freidel:1991jx,Freidel:1991jv}, and therefore we omit tedious computational details. 

Finally, invoking the relation between Maillet's symmetrized brackets and the quantum commutator regularised by means of the Sklyanin product \eqref{QA:SP_sym_connection}, we can accordingly conclude that in the classical limit the quantum algebra given by \eqref{QA:equal_interval}  and \eqref{QA:adjacent_interval} reduces to the classical algebra \eqref{sp:Maillet_classical_algebra}. Thus, we have shown how to obtain the classical algebra \eqref{sp:Maillet_classical_algebra} in terms of Maillet's symmetrized Poisson brackets from the quantum relations \eqref{QA:equal_interval} and \eqref{QA:adjacent_interval} involving commutators with respect to Sklyanin product. This result allows us to interpret Maillet's symmetrization prescription for Poisson brackets \eqref{intro:nested_brackets}, \eqref{intro:symmetrization} simply as the consequence of the regularisation of the singular operator product at the same point in the quantum case, when taking the classical limit. 

To summarise, instead of starting from the classical Poisson brackets, where in order to avoid some problems with, e.g., the Jacobi identity, one has to modify \emph{by hand} the definitions and utilize the symmetrization procedure of Maillet, it is possible to start from  the more natural standpoint of the quantum algebra where singular operator products are regularised, and then obtain the classical algebra by taking the classical limit. As a final remark, we would like to stress that the usage of Sklyanin product in the previous expressions is not merely formal. For example, in the case of the anisotropic Heisenberg model, the quantum algebra is formulated in terms of the Sklyanin product \cite{Sklyanin:1988}, which makes it possible to extract the conserved commuting quantities and construct the quantum states. 

\section{Conclusion}\label{sec:conclusion}

In this short note we showed that Maillet's symmetrization prescription for classical Poisson brackets can be naturally obtained from a quantum algebra, where all operator products are regulalised by means of Sklyanin product. The latter essentially corresponds to a ``smeared'' product of operator-valued functions over a small neighbourhood around the singular region. Furthermore, we have also established that the Yang-Baxter-like equations constraining the classical quadratic algebra of Freidel-Maillet type follow from the quantum Jacobi identity in the classical limit.

The analysis in this paper was restricted to the particular case of bosonic fields and constant numerical matrices $a_{12}$, $b_{12}$, $c_{12}$ and $d_{12}$ encoding the classical Maillet-Freidel quadratic algebra, which depended nonetheless on the spectral parameters. However, it is well known that for many interesting non-ultralocal integrable models such matrices are coordinate dependent as well. Hence, one needs to generalize the results obtained here for such coordinate dependence of the parameters of the quadratic algebra. 

As in the case of Maillet's symmetrization procedure for $n$-nested brackets, the regularisation of the quantum algebra in terms of the Sklyanin product we employed in this paper  requires a multi-step prescription. Namely, it is defined for each $n$-operator product as in equation \eqref{sp:sklyanin_prod_general}. For the case of Maillet brackets it was demonstrated in \cite{Forger:1991ty} that a single-step regularisation, which would reduce the definition of a nested Poisson bracket to repeated application of a basic regularised Poisson bracket involving only two fields, is incompatible with the Jacobi identity. Therefore, it is an interesting open problem to consider whether this limitation also applies to our quantum regularisation.

One should solve similar questions, as well as the problem of formulating a well-defined quantum algebra of transition matrices, according to the principles of axiomatic quantum field theory, consistently defining the product of operator-valued distributions. Out of several different approaches to evade Schwartz's impossibility theorem, two look specially promising: H\"ormander's wavefront set of distributions and Colombeau algebras. The former has been used in the context of quantum field theory, making it possible to define the product of distributions. However, even in this approach some interesting products of distributions, e.g., the powers of the Dirac delta distribution, remain ill-defined.  As for the latter, the space of Schwartz distributions is embedded into an associative algebra which satisfies the Leibniz's rule, nevertheless the association between a distribution and an element of the Colombeau algebra is not always unique. Thus, it is still an open and interesting question whether any of these approaches will prove to be successful in the context of the quantisation of continuous non-ultralocal integrable systems. We also mention here that another compelling possibility to approach this problem lies within Sato's theory of hyperfunctions.

\section*{Acknowledgments} G.W. would like to thank D. Guariento for useful discussions. The work of A.M. is partially supported by CAPES. 
\phantomsection
\addcontentsline{toc}{section}{References}
\bibliographystyle{utphys}
\bibliography{references}

\end{document}